\newcommand{\LTO}{\text{LaTiO}_\text{3}} 
\newcommand{\STO}{\text{SrTiO}_\text{3}} 
\newcommand{\KTO}{\text{KTaO}_\text{3}} 
\newcommand{\LAO}{\text{LaAlO}_\text{3}} 
\newcommand{\Rxx}{R_\text{xx}} 
\newcommand{\Tconset}{T_\text{c}^{\text{onset}}} 

\documentclass[aip,graphicx,amsmath,amssymb,reprint,twocolumn]{revtex4-2}
\usepackage{graphicx}
\usepackage{dcolumn}
\usepackage{bm}
\usepackage{epsfig}
\usepackage{natbib}
\usepackage{color}
\usepackage{xcolor}
\usepackage[utf8]{inputenc}
\usepackage{graphicx}
\usepackage{dcolumn}
\usepackage{bm}
\usepackage{ulem}

\usepackage{mathptmx}
\usepackage{etoolbox}

\begin{document}


\title{Superconductivity at epitaxial LaTiO$_3$-KTaO$_3$ interfaces} 

\author{D. Maryenko}
\email{maryenko@riken.jp}
\affiliation{RIKEN Center for Emergent Matter Science (CEMS), Wako 351-0198, Japan}
\author{I. V. Maznichenko}
\affiliation{Institute of Physics, Martin Luther University Halle-Wittenberg, 06120 Halle, Germany}
\author{S. Ostanin} 
\affiliation{Institute of Physics, Martin Luther University Halle-Wittenberg, 06120 Halle, Germany}
\author{M.~Kawamura}
\affiliation{RIKEN Center for Emergent Matter Science (CEMS), Wako 351-0198, Japan}
\author{K.~S.~Takahashi}
\affiliation{RIKEN Center for Emergent Matter Science (CEMS), Wako 351-0198, Japan}
\author{M.~Nakamura}
\affiliation{RIKEN Center for Emergent Matter Science (CEMS), Wako 351-0198, Japan}
\author{V.~K.~Dugaev}
\affiliation{Department of Physics and Medical Engineering, Rzesz\'{o}w University of Technology, 35-959 Rzesz\'{o}w, Poland}
\author{E.~Ya.~Sherman}
\affiliation{Department of Physical Chemistry and EHU Quantum Center, University of the Basque Country, 48940, Leioa, Spain} 
\affiliation{Ikerbasque, Basque Foundation for Science, Bilbao, Spain}
\author{A.~Ernst}
\affiliation{Institute for Theoretical Physics, Johannes Kepler University, 4040 Linz, Austria}
\affiliation{Max Planck Institute of Microstructure Physics, D-06120 Halle, Germany}
\author{M. Kawasaki}
\affiliation{RIKEN Center for Emergent Matter Science (CEMS), Wako 351-0198, Japan}
\affiliation{Department of Applied Physics and Quantum-Phase Electronics Center (QPEC), The University of Tokyo, Tokyo 113-8656, Japan}

\date{\today}

\begin{abstract}
Design of epitaxial interfaces is a pivotal way to engineer artificial structures where new electronic phases can emerge. Here we report a systematic emergence of interfacial superconducting state in epitaxial heterostructures of $\LTO$ and $\KTO$. The superconductivity transition temperature increases with  decreasing the thickness of $\LTO$. Such behavior is observed for both (110) and (111) crystal oriented structures. For thick samples, the finite resistance developing below the superconducting transition temperature increases with increasing $\LTO$ thickness. Consistent with previous reports, the (001) oriented heterointerface features high electron mobility of 250 cm$^2$V$^{-1}$s$^{-1}$ and shows no superconducting transition down to 40 mK. Our results imply a non-trivial impact of $\LTO$ on the superconducting state and indicate how superconducting $\KTO$ interfaces can be integrated with other oxide materials.
\end{abstract}

\pacs{}

\maketitle 

\section*{Introduction}
Interfaces between materials can harbor electronic structures distinct from the bulk constituents. One instance is the formation of a metallic layer at the junction of two insulators. A broadly celebrated example is $\LAO /\STO$ interface, that harbors not only high mobility carriers but can also become superconducting at around 300~mK~\cite{HaroldLAOSTO, LAOSTOSuperconductance1,LAOSTOSuperconductance2}.  This rather well controlled system became a fertile testbed to explore two-dimensional superconductivity. In such a strongly assymetric heterostructure, it was straightforward to assay the role of spin-orbit coupling (SOC) for the superconducting phase, albeit the conduction band is formed by $3d$-orbitals of titanium with a moderate SOC energy  on the order of 40~meV~\cite{Rashba,CavigliaSOC,DaganSOC,HerranzSOC}. In fact, it is anticipated that a sizable spin-orbit coupling can be favorable for unconventional Cooper pairing  and for realization of Majorana states ~\cite{VenderbosPRX2018, GorkovRashbaSC, UnconventionalSC1,UnconventionalSC2,UnconventionalSC3}. Therefore the recent observation of superconductivity in $\KTO$, whose conduction band is formed by $5d$ Ta orbitals with  a much larger SOC energy of about 300~meV,  may provide a new twist in the formation of superconducting phase in two dimensions. Furthermore, by taking into consideration that bulk $\KTO$ has not still been demonstrated to become superconducting, the emergence of interfacial superconductivity  in such a system can  provide a distinct insight into Cooper pair formation mechanism~\cite{KTOWemple}.  Being isostructural to $\STO$, perovskite oxide $\KTO$ is a quantum paraelectric and has a band gap of about 3.6~eV. The conduction band around $\Gamma$ point is split  by a large spin-orbit coupling in well separated bands  with an effective total angular momentum $J=1/2$ (higher energy) and $J=3/2$ states (lower energy). 

The first observation of interfacial $\KTO$ superconductivity dates back to experiments with the ionic liquid gating technique, that has revealed a superconducting transition at 50~mK for (001)-oriented $\KTO$ surface~\cite{KTOUeno}. Recently the emergence of superconductivity  at (110)- and (111)-oriented $\KTO$ surfaces is demonstrated in the majority of cases by growing a EuO layer or depositing amorphous $\LAO$ layer ~\cite{LAOKTO,EuOKTOScience,EuOKTO110, LAOKTOScience, EuOKTONatComm, AlOKTO,TiOKTO}.  
The cubic lattice structure of EuO with a lattice constant $a=5.145~\text{\AA}$  matches neither (110) nor (111) orientation of $\KTO$ crystal  structure, resulting in the formation of either polycrystalline or defective layers at the interface~\cite{EuOKTOScience, EuOKTO110}.  Superconductivity was also observed in (111)-oriented $\KTO$ heterostructure with a 10 nm thick La$_{2/3}$Sr$_{1/3}$MnO$_3$ top layer \cite{KTOScienceAdvances} . 
To have full control over the emergent superconducting state, it is important  to have excellent control over the interface's electronic properties, which also includes understanding of the role of the top layer for the emergent phenomena. This control paves the way for integrating superconducting $\KTO$ interfaces with other oxide materials. 

 \begin{figure*}
 \includegraphics{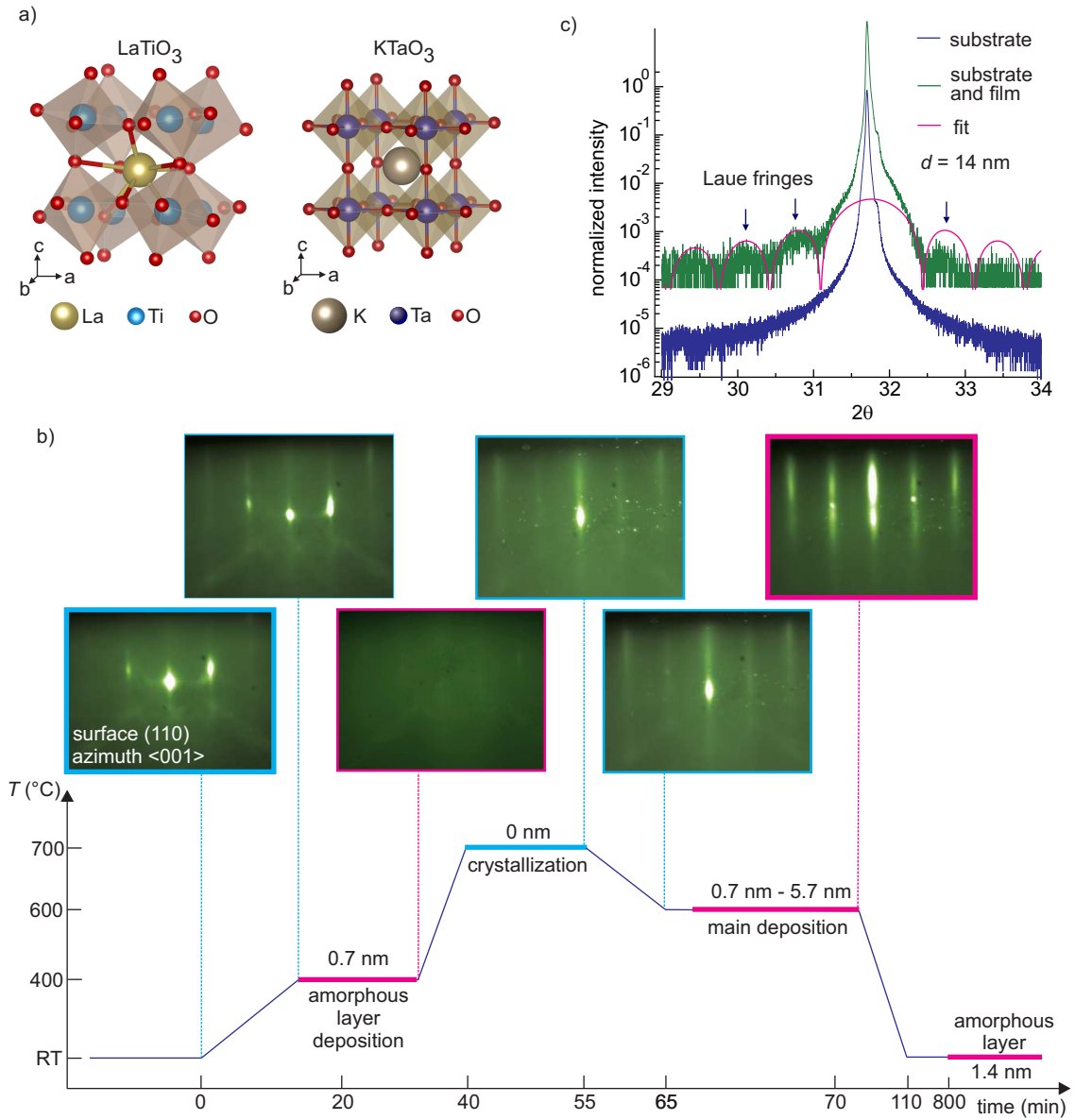}
 \caption{\label{Fig1} a) Crystal structure of $\LTO$ and $\KTO$ \cite{Drawing}. b)  Epitaxial growth process steps for $\LTO/\KTO$ heterostructures. Shown are RHEED patterns at various steps of (110) oriented structure growth. Similar evolution of RHEED pattern with temperature is also observed for structures grown on (001) and (111) $\KTO$ crystal orientations. c) X-ray diffraction patterns (110)-oriented substrate (blue trace) and film on substrate(green trace). The diffraction pattern are shifted for clarity along  vertical axis. Red line is the best fit describing the position of Laue fringes. }
 \end{figure*}

Here, we report the emergence of superconductivity in epitaxial grown structures of $\LTO$ on (110) and (111) oriented $\KTO$.  We observe that the superconducting transition temperature increases with decreasing thickness of the $\LTO$ layer.  For thick samples, the resistance $\Rxx$ remains finite below superconducting transition temperature and this $\Rxx$ value increases with increasing $\LTO$ thickness. These observations indicate a non-trivial impact of $\LTO$ on the interface's electronic properties. Our finding may facilitate engineering of the superconducting phase at the interface. Bulk $\LTO$ is a Mott insulator with orthorhombic crystal structure and lattice parameters $a=b=5.595~\text{\AA}$ and $c=7.912~\text{\AA}$.  Therefore, $\LTO$ can be thought as a quasi cubic structure with an effective lattice constant $\sqrt{a^2+b^2}/2 \cong c/2= 3.956~\text{\AA}$, which thus differs by only about 0.1$\%$ from the lattice constant of cubic $\KTO$ $a=b=c= 3.989~\text{\AA}$.  This facilitates the growth of $\LTO/\KTO$heterostructure on the three main facets of a cubic crystal system, i.e. (001), (110), and (111)\cite{LTOKTOAhn}.

  \begin{figure*}
 \includegraphics{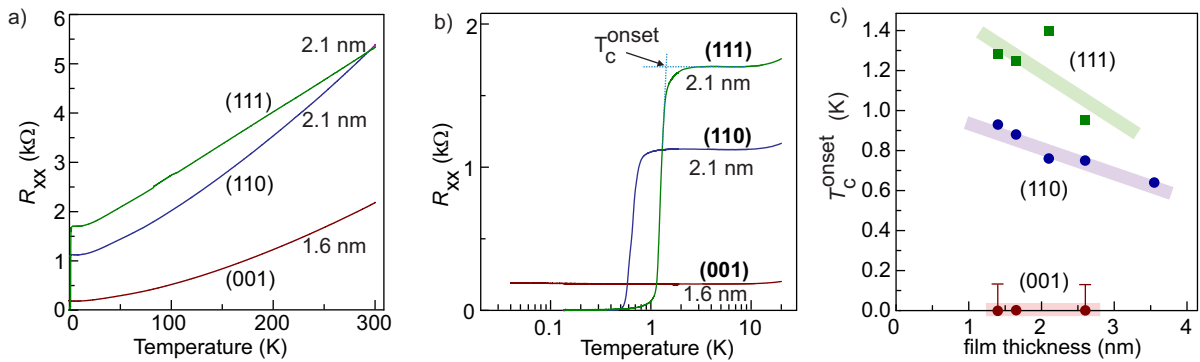}
 \caption{\label{Fig2} a) Exemplary temperature dependence of resistance for $\LTO/\KTO$ heterostructures defined on (001), (110) and (111) crystal surfaces. b) Superconducting state is observed for (110) and (111) oriented heterostructures, while (001) structure remains metallic down to 40~mK. Shown is the definition of the superconductivity onset temperature $\Tconset$.  c) $\Tconset$ decreases with increasing the thickness of $\LTO$. We assign an error bar of 150~mK for (001)-oriented heterostructures, that are not measured in dilution refrigerator. Thick lines are guides to the eye.}
 \end{figure*}

\section*{Results and Discussion}
\subsection{Epitaxial growth}
$\LTO/\KTO$ structures are grown using pulsed laser deposition technique. A piece of $\KTO$ substrate with a size of about 3~mm x 3~mm was attached to the substrate holder using silver epoxy. A polycrystalline La$_2$Ti$_2$O$_7$  target is ablated in vacuum with a repetition rate of 2~Hz and a laser fluence 1.6~Jcm$^{-2}$. The growth chamber is equipped with a reflection high-energy electron diffraction (RHEED) monitor allowing us to observe the growth process \textit{in-situ}. Figure~1b depicts exemplary RHEED patterns during the growth process of (110)-oriented structure. After loading the substrate in the growth chamber,  the substrate is heated to 400$^\circ$C. During this heating step no change in the RHEED pattern is observed. In fact, the atomic force microscopy measurements show that the surface morphology barely changes at 400$^\circ$C (see Supplementary Information). To prevent the degradation of $\KTO$ surface upon further heating and to suppress the formation of defects, the substrate surface is covered with an amorphous layer by ablating La$_2$Ti$_2$O$_7$ target, which is indicated by the vanishing RHEED pattern after this process step.  Upon heating to 700$^\circ$C the amorphous layer crystallizes and the streak pattern forms gradually. This solid state epitaxial step at 700$^\circ$C is favored due to a small lattice mismatch between $\LTO$ and $\KTO$, which gives a clear diffraction pattern correspondence between the substrate and the crystallized layer. The crystallized layer enables successive homoepitaxial growth, which  takes place at a lower temperature of 600$^\circ$C. The heterostructures discussed in this work differ by the $\LTO$ layer thickness deposited at 600$^\circ$C. After the growth, the heterostructures are cooled to room temperature and are left to thermalize for about 12h. Subsequently, the structures are covered with a thin amorphous layer by ablating the La$_2$Ti$_2$O$_7$ target to prevent  potential degradation of structures at ambient condition. We note that the growth conditions favor the stabilization of $\LTO$ phase \cite{Ohtomo2002}.  To check the film crystal structure, we grow a thick $\LTO$ layer with 735 pulses. Figure 1c depicts its x-ray diffraction pattern (green trace) featuring  Laue fringes, which indicate a high crystalline film quality. Due to the similar lattice parameters of $\LTO$ and $\KTO$, the Bragg diffraction peak of $\LTO$ is indiscernible due to the overlapped diffraction pattern of the substrate (blue trace). By fitting the position of the Laue fringes (red line) we determine the film thickness of 14~nm,  which is used to estimate the thickness of thinner films from a given number of pulses as depicted in Fig.~1b for each process step. The stoichiometry of the structures checked with energy-dispersive X-ray spectroscopy was comparable with that of the target. We note however that the exact stoichiometry of the structure can have an impact on the interface conductivity~\cite{LAOSTOStoichiometry1, LAOSTOStoichiometry2, LAOSTOStoichiometry3}.

\subsection{Electrical Transport Characteristics}
The transport characteristics of heterostructures are shown in Fig.~2.  We employed a Physics Properties Measurement System (PPMS, Quantum Design) down to 2~K and adiabatic demagnetization refrigerator (ADR) stage, which is compatible with PPMS platform, to characterize the superconducting transition of heterostructures down to 150~mK. The samples are directly bonded with aluminum wires as depicted in Fig. 3a, so that they can be characterized along two orthogonal directions simultaneously.  
Figures 2a and 2b depict exemplary temperature dependence of $\Rxx$ for three crystal orientations. Consistent with previous reports, the (111)-oriented heterostructure has a higher superconducting transition temperature than the (110)-oriented heterostructures~\cite{LAOKTO,EuOKTOScience, EuOKTO110, EuOKTONatComm}.  More importantly, we observe that the onset temperature of superconducting phase $\Tconset$ strongly depends on the thickness of $\LTO$ layer. Such an impact of top layer thickness on the superconducting state hasn't been reported yet. Figure 2c depicts that $\Tconset$ increases with decreasing thickness of the $\LTO$ layer. Such behavior is observed for both (110) and (111) oriented heterostructures. Following this finding, we measured one of the (001)-oriented heterostructures with a 1.7~nm thick $\LTO$ layer in a dilution refrigerator at temperature down to 40~mK, but did not observe superconducting transition. The absence of superconducting phase for (001) oriented heterostructures is consistent with previous report~\cite{EuOKTONatComm}. To check  the conductance of $\LTO$ layer, we grew a 2.6~nm thin $\LTO$ layer on both GdScO$_3$ and NdScO$_3$ substrates according to the growth procedure of Fig.~1. The resistance of such structures at room temperature was on the order of 10$^6$~Ohm.

 \begin{figure*}
 \includegraphics{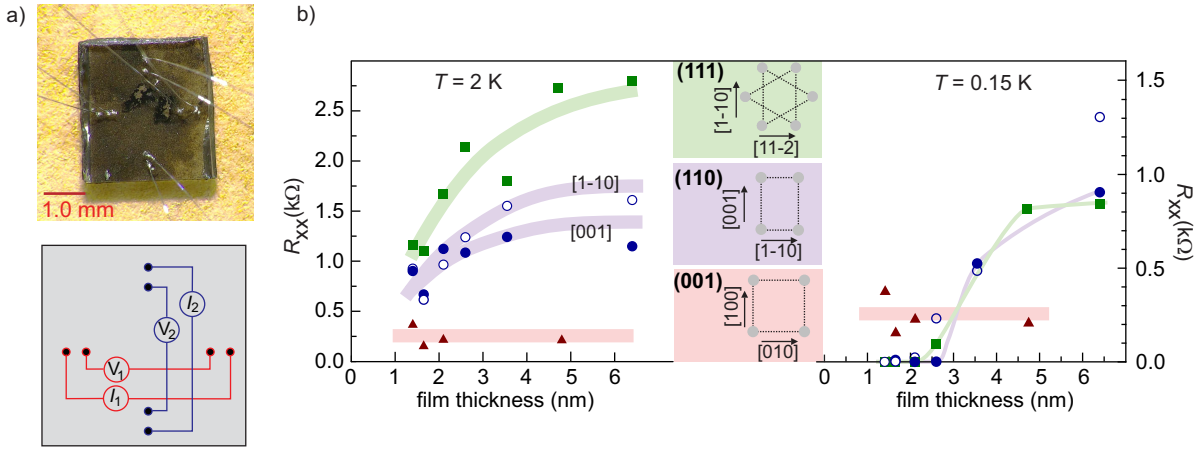}
 \caption{\label{Fig3}a) Photograph of a sample with attached wires to measure temperature dependence of sample resistance. Shown also is the scheme of electrical circuit sample connection. Multi-channel source-measurement unit of PPMS is used to measure two orthogonal crystal directions simultaneously. b) Resistance at zero magnetic field at $T=$2~K (above superconducting transition, left panel) and at $T=$~150~mK (below superconducting transition, right panel) as a function of $\LTO$ thickness. The color encodes the crystal orientation of heterostructures. Thin heterostructures with (110) and (111) crystal orientations show $\Rxx=0~\Omega$ at $T=$~150~mK. As $\LTO$ thickness increases, the residual $\Rxx$ increases. The thick lines are guides to the eye. }
 \end{figure*}
Figure 3b compares the resistance values of heterostructures above ($T=$~2~K, left panel) and below ($T=$~150~mK, right panel) the superconducting transition. It is  noticeable that $\Rxx$ at $T=$~2~K increases with increasing $\LTO$ thickness for both the (111) and (110) oriented structures, while it remains almost constant for the (001)-oriented heterostructures.  Such behavior points to the interface conductance rather than to the conductance in $\LTO$ layer solely, in which case the resistance would decrease with increasing $\LTO$ thickness. To further elucidate the properties of the heterostructures, we present in Fig. 4 the dependence of both the electron mobility and the charge carrier density on $\LTO$ thickness. Charge carrier density $n$ is estimated from the Hall effect measurements,  while mobility $\mu$ is estimated from the sample conductance in zero magnetic field. Among the three crystal orientations, the (001)-oriented heterointerface has the highest electron mobility on the order of 250 cm$^2$V$^{-1}$s$^{-1}$, which does not depend on the $\LTO$ thickness. Both the electron mobility and the charge carrier density values are consistent with those obtained for $\LTO/\KTO$ (001)-oriented structures grown by molecular beam epitaxy ~\cite{LTOKTOAhn}.  For both the (110) and  (111) oriented heterostructures the electron mobility shows a distinct behavior; it is the largest for thin structures, i.e. around 100 cm$^2$V$^{-1}$s$^{-1}$, and decreases with increasing $\LTO$ thickness, reaching a saturation value of around 30~cm$^2$V$^{-1}$s$^{-1}$ above a $\LTO$ thickness of 2~nm.  By contrast the charge carrier density shows a fast increase with $\LTO$ thickness by about a factor 1.5 (lower panel in Fig.~4) and saturates above 2~nm. This seems to be a common tendency for all three crystal orientations. An increase of the sheet charge carrier density $n$ (Fig.~4b) and, at the same time, a decrease of $\Tconset$ with $\LTO$ thickness (Fig.~2c) establishes an opposite tendency to the previous  observations in $\KTO$-based superconducting structures, for which superconducting transition temperature increases with increasing $n$~\cite{EuOKTONatComm, AlOKTO}. The presented results indicate an impact of epitaxial $\LTO$ layer on the electronic properties of the interface, which also affects the superconducting regime as we discuss now.   

 \begin{figure}
 \includegraphics{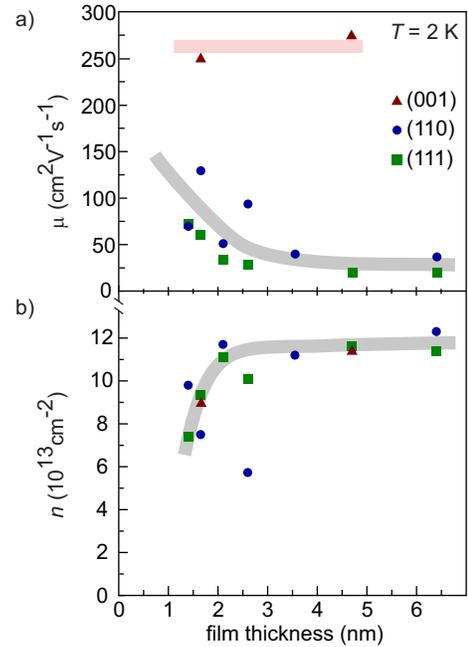}
 \caption{\label{Fig4} Mobility (panel a))  and charge carrier density (panel b))  dependence on $\LTO$ layer thickness at $T =$ 2~K for different heterostructure orientations. The charger carrier density is estimated from the transverse resistance $R_\text{xy}$ (Hall effect), which changes linearly with the magnetic field $B$. Thick lines are guides to the eye.}
 \end{figure}

Right panel in Fig.~3b depicts the dependence of $\Rxx$ at 150~mK (well below $\Tconset$) on  the $\LTO$ thickness for all heterostructures. For the sake of comparison, it also contains data points for the high mobility (001) interface that does not become superconducting in our experiments. A well developed superconducting state characterized by  $\Rxx=0~\Omega$ is  reached  for both (110)- and (111)-oriented heterostructures but only with a thin $\LTO$ layer. For thicker $\LTO$ layers $\Rxx$ attains a non-zero value, which increases with increasing $\LTO$ thickness.  
Furthermore, we detect an anisotropy for the (110)-oriented structures. Open symbols in the right panel depict $\Rxx$ values measured along the [1-10] direction at 150~mK. 
For 2.1~nm and 2.6~nm thick samples the superconductivity along [001] direction survives at 150~mK, while $\Rxx$ along [1-10] direction has a non-zero value.
 (In the Supplementary Information we show the temperature dependence of $\Rxx$ during  superconducting transition for all samples.)
By contrast to (110)-oriented structures,  [1-10] and [11-2] crystal directions of (111)-oriented heterostructures appear to be equivalent. Since the heterostructures are grown in equivalent procedures, it allows us to conclude that a potential sample inhomogeneity cannot explain the anisotropy  as observed in (110)-oriented structures. This surprising emergence of anisotropic behavior of $\Rxx$ below superconducting transition is perhaps related to the immanent electronic structure of the interface. 
Anisotropy for (110)-oriented heterostructures is reported for the normal conducting state of $\STO$-based heterostructures and is related to a different arrangement of interface atoms along [001] and [1-10] directions~\cite{LAOSTOAnisotropy,STOAnisotropy}. 
An indication of such an anisotropy in our (110)-oriented heterostructures might also appear in the normal conducting state. At $T=$ 2~K, Figure~3b (left panel) depicts that $\Rxx$ along the [1-10] direction (open blue symbols) is larger than for the [001] direction (full blue symbols). 

Beyond that, increasing the $\LTO$ thickness affects the transport characteristics of both the (110)- and (111)-oriented heterostructures before superconducting transition. In fact, for structures with a thicker $\LTO$, one clearly observes some increase in  $\Rxx \propto \ln T$  indicating a contribution of the weak localization correction to the sample resistance (see Supplementary Information). This has also been observed in superconducting $\LTO/\STO$ structures \cite{LTOSTONatComm}. Conspicuously, when this localization behavior is strongly pronounced in our structures, $\Rxx=0~\Omega$ vanishes for (110) as well as for (111)-oriented structures, as seen in Supplementary Information. The (110) and (111) heterostructures feature a weak antilocalization behavior in magnetotransport, indicating a significance of spin-orbit coupling.  Intriguingly, weak antilocalization is barely pronounced for non-superconducting (001)-oriented heterostructures (see Supplementary Information).

The observation of the superconducting transition being dependent on the $\KTO$ surface orientation is consistent with previous reports on superconductivity in $\KTO$\cite{LAOKTO, EuOKTOScience, EuOKTO110, LAOKTOScience, LAOKTO, EuOKTONatComm}. This allows us to conclude that the superconducting phase in our structures involves the electronic states of $\KTO$. At the same time the $\LTO$ thickness dependence of the transport characteristics both in superconducting and normal state implies a non-trivial impact of the top layer on the electronic structure of the $\LTO/\KTO$ heterointerface. In the vicinity of the junction, the Ta atoms are in a 5+ state, whereas Ti is in a 3+ state. This charge discontinuity can lead to charge redistribution between the $\LTO$ and $\KTO$ layers adjacent to the interface, creating an interfacial conducting layer. One such mechanisms can be related to a so called polar catastrophe, which is based on compensation of the diverging electrostatic energy at the interface~\cite{PolarCatastrophe}. This mechanism has been considered for various $\STO$ and $\KTO$  based heterostructures and can be effective for (001) and (111) oriented structures, but is not obvious for (110) structures ~\cite{LTOSTO, LVOSTO, GTOSTO, LTOKTOAhn, LCOKTO}. Surface reconstruction and the modification of TiO$_6$ octahedra have also been considered for the emergence of conducting layers at the interface between band insulators and Mott insulators such as $\LTO$~\cite{OkamotoMillis, LTOSTOSpaldin, LTOSTOMetallicity}. Moreover, oxygen defects can contribute to the emergence of a conducting layer. It would require additional experimental and theoretical efforts to elucidate how each of those mechanisms is realized in our superconducting $\LTO/\KTO$ structures. The interplay of those mechanisms will define the extension of the conducting layer,  interaction between the $\LTO$ and $\KTO$ layers, and consequently the total electronic structure. 

\section*{Conclusion}
In summary, we have grown epitaxial $\LTO/\KTO$ heterostructures with (001), (110) and (111) crystal orientations and varying $\LTO$ thickness. The (110)- and (111)-oriented heterostructures have a moderate electron mobility and a well developed superconducting state. The (001)-oriented heterostructures have the highest electron mobility with no indication of a superconducting transition. The $\LTO$ layer has a non-trivial impact on the emergence of the superconducting phase. With increasing $\LTO$ thickness the superconducting transition temperature decreases and a finite resistance remains below the transition. This behavior seems to correlate with the emergence of electron weak localization.  Furthermore, for the (110)-oriented heterostructures we observe a regime when $\Rxx = 0~\Omega$ along the [001] direction and non-zero for the [1-10] direction, thus establishing anisotropic superconductivity  in $\LTO/\KTO$-heterostructures. Our result may pave the way to engineer superconducting interfaces and to integrate superconducting $\KTO$ interfaces with oxide materials. 

\section*{Supplementary Materials}
See the supplementary material for additional details on superconducting transition, features of weak antilocalization behavior in the magnetic field and atomic force microscopy images.  

\begin{acknowledgments}
We would like to thank Dr. M. Kriener and Dr. M. Birch for fruitful discussion and careful reading of the manuscript. This work was supported by JSPS KAKENHI (22H04958). The work of E.S. is supported through Grants No. PGC2018-101355-B-I00 and No. PID2021-126273NB-I00 funded by MCIN/AEI/10.13039/501100011033 and by the ERDF “A way of making Europe”, and by the Basque Government through Grant No. IT1470-22. The work of V.D. is supported by the National Science Center in Poland as a research project No. DEC-2017/27/B/ST3/02881.

\end{acknowledgments}

\nocite{*}

\end{document}